\documentclass[prd,11pt,aps,superscriptaddress,floatfix,showkeys,showpacs]{revtex4}
\usepackage{graphics,epsfig}
\usepackage{graphicx,amsmath,bm,array,subfigure}
\usepackage[linktocpage=true,colorlinks=true,linkcolor=blue,citecolor=blue]{hyperref}
\usepackage{lmodern}
\usepackage {amssymb}

\begin{document}
\title {Probing vorticity in heavy ion collision with dilepton production}
\author{Balbeer Singh }
\email{balbeer@prl.res.in}
\affiliation{Theory Division, Physical Research Laboratory,
Navrangpura, Ahmedabad 380 009, India}
\affiliation{Indian Institute of Technology Gandhinagar, Gandhinagar 382355, Gujarat, India}
\author{Jitesh R Bhatt}
\email{jeet@prl.res.in}
\affiliation{Theory Division, Physical Research Laboratory,
Navrangpura, Ahmedabad 380 009, India}
\author{Hiranmaya Mishra }
\email{hm@prl.res.in}
\affiliation{Theory Division, Physical Research Laboratory,
Navrangpura, Ahmedabad 380 009, India}
\begin{abstract}
We study the effect of vorticity present in heavy ion collisions (HICs) on the temperature  evolution of hot quark gluon plasma in the presence of spin-vorticity coupling. The initial global rotation entails a nontrivial dependence  of the longitudinal flow velocity on the transverse coordinates and also develops a transverse velocity component that depends upon the longitudinal coordinate. Both of these velocities lead to a 2+1 dimensional expansion  of the fireball. It is observed that with finite vorticity and spin-polarization the  fireball cools faster as compared to the case without vorticity. Furthermore, we discuss the consequence of this on the production of thermal dileptons.
\end{abstract}


\maketitle

\section{Introduction}

One of the most intriguing features of matter created in heavy-ion experiments at LHC and RHIC is the nearly ideal fluid behavior.
These experiments create a phase of strongly interacting matter called quark-gluon plasma (QGP). 
Here the fluid dynamical behavior sets in at very early stages during the collision due to large enough interaction rates among the partons. 
Relativistic hydrodynamics with minimal viscous corrections has become an extremely useful tool in describing these collision experiments ~\cite{Gale:2013da,Heinz:2013th}.
For example the hydrodynamic models, with the ratio of  shear viscosity  $\eta$ to entropy density $s$: $\eta/s < 0.2$, have been quite successful in describing the elliptic flow data. 
Such a small value of  $\eta/s$ is closed to the conjectured lower bound $\eta/s\sim 1/4 \pi$ which
is known as the Kovtun-Son-Starinets (KSS) bound. This bound is the quantum limit suggested by anti-de Sitter/conformal field theory (AdS/CFT) correspondence ~\cite{Kovtun:2004de}. This qualifies the strongly interacting matter created in RHIC experiments as one of {\it the most perfect fluid}.

Another puzzling aspect of strongly interacting matter created in heavy-ion collisions is  polarization of  $\Lambda$ hyperon implying very high fluid vorticity. For example, for a non-central collision the total initial angular momentum $L_0 \propto b \sqrt{s}$, where, $b$ is impact parameter and $\sqrt{s}$ denote the center of mass energy.
Thus in an Au-Au  collision at RHIC energies with $\sqrt{s}=200$ GeV/nucleon one estimates $L_{0} \sim  10^5$.  For
Pb-Pb collisions at LHC energies with $\sqrt{s}=5.5$ TeV/nucleon, $L_{0}\sim 10^7$ for $b=5$ fm ~\cite{Becattini:2007sr}. A fraction of this initial angular momentum could be retained in the interaction region which will eventually be transferred to the QGP. This fraction of $L_0$ can manifest itself as a shear of the longitudinal momentum density. Thus in a direction perpendicular to the reaction plane, non-zero local vorticity is created. 
As a signature of the vorticity in the fluid created in heavy-ion  collisions, the polarization of emitted hadrons was suggested. Indeed, the study of the polarization of $\Lambda$ hyperon leads to an estimate of  the vorticity $\omega \sim (9 \pm 1) \times 10^{21}$  ${s}^{-1}$\cite{STAR:2017ckg}. This also makes the matter created in the heavy-ion collision as  {\it the perfect vortical fluid}. One of the important observations that came out of this study is that the event averaged vorticity decreases with an increase in the collision energy \cite{STAR:2017ckg}.

There are several efforts to study the generation and the dynamics of vorticity in the context of heavy-ion collision experiments. In Ref.~\cite{Jiang:2016woz}, using a multiphase transport(AMPT) model, the global angular momentum and vorticity carried by QGP have been estimated for Au-Au collisions. Rotation and vorticity in  peripheral heavy-ion collisions within the hydrodynamics framework have been studied in ~\cite{Gao:2014coa,Csernai:2014hva}. Further, in Ref \cite{Deng:2016gyh}, it has been shown that for large Reynolds number $Re=u L \eta^{-1}$, which at RHIC energy scale varies between $10-100$ for flow velocity $u=(0.1-1)$, fluid length scale $L=5 $ fm and temperature $T=300$ MeV, the vortex lines are frozen to the fluid and the vorticity decreases owing to the expansion of the system. However, for smaller Reynolds number, the vorticity is damped and rather decreases rapidly.
Further, it is worth mentioning here that in the presence of non-vanishing vorticity, the longitudinal velocity develops a dependence on the transverse coordinate. It is therefore expected that, with finite vorticity, the hydrodynamic expansion of the thermalized plasma will be very different from the usual 1D Bjorken flow. This makes the hydrodynamic expansion in the presence of vorticity to be in 2+1 dimensions.  

Consequences of finite vorticity have been investigated in a variety of situations in LHC and RHIC experiments.
One of the manifestations  of non-zero vorticity lies in the polarization of the secondary-particles \cite{Liang:2004ph, Betz:2007kg, Karpenko:2016jyx,Xie:2016fjj, STAR:2017ckg,Upsal:2016phr}, 
Another interesting effect is that of the quark and anti-quark global polarization due to the spin-orbit coupling \cite{Liang:2004ph}. This leads to observable effects like emission of circularly polarized photons \cite{Ipp:2007ng} and spin-alignment of vector mesons  \cite{Liang:2004xn}. A  two-particle correlation function has also been proposed in Ref.\cite{Csernai:2013vda} by employing differential Hanbury Brown and Twiss (HBT) analysis. It may also be possible to have parity odd effects due to vorticity associated with the chiral vortical effect (CVE) leading to charge separation and induced currents \cite{Kharzeev:2007tn, Rogachevsky:2010ys} as well as chiral vortical wave (CVW). The latter one leads to the elliptical flow splitting of baryons and anti-baryons \cite{Jiang:2015cva}.
 
In this work, we consider the effect of spin-vorticity coupling on the thermal evolution of the rotating fluid created in HIC experiments. The spin-vorticity coupling, responsible for the Bernnet effect ~\cite{Bernett:2015, Fukushima:2018osn}, can possibly modify the thermodynamic relation as \cite{Becattini:2009wh,Florkowski:2017ruc}
\begin{equation}
\epsilon+P=Ts+\mu n+\Omega w,
\label{eos1}
\end{equation}
where, $\epsilon$, $P$, $T$, $s$, $\mu$ $n$ respectively denote energy density, pressure, temperature, entropy density, chemical potential, number density. In the last term of Eq.[\ref{eos1}], $\Omega$ is the vorticity term defined as $\Omega=\frac{T}{2\sqrt{2}} \sqrt{\omega^{\mu \nu}\omega_{\mu \nu}}$ where $\omega_{\mu \nu}$ is the spin polarization tensor. $\Omega$ can also be regarded as the spin chemical potential corresponding to spin density $w$ ~\cite{Florkowski:2017ruc}. For a system in thermodynamic equilibrium, $\Omega$ is proportional to thermal vorticity \cite{Becattini:2013fla}.
Here it should be noted that for local thermodynamic equilibrium this assertion may require a careful assessment~\cite{Becattini:2018duy}. It ought to be noted that Eq.[\ref{eos1}] is obtained for a specific choice of the energy-momentum tensor along with a phenomenological spin tensor\cite{Florkowski:2017ruc,Florkowski:2018fap} which is assumed to be conserved. 
In a careful analysis, it was shown in Ref.\cite{Florkowski:2018fap} that the choice for the conserved energy-momentum  and  spin tensors are not unique. In particular, it was shown that the different choices of energy-momentum and spin tensors are related by a pseudo gauge transformation. For the present investigation, however, we consider that spin degree of freedom has been equilibrated and the spin as a hydrodynamical variable is not important. We would like to emphasize that Eq.(\ref{eos1})  will have a contribution from spin-orbit coupling as it remains invariant under the pseudo-gauge transformation. The use of Eq.(\ref{eos1}), we believe, will provide useful insight about the nature of the thermodynamic relation  in the presence of spin-polarization. Further, we shall restrict ourselves to the limit of large Reynolds numbers only. Thus the effect of viscosity will be neglected in the present analysis.

At present, there have been both theoretical and experimental efforts to understand the role of large vorticity in the collision experiments. Phenomenologically, it is also important to quantify the rotational motion of the QGP in these collisions. 
We show that the spin-vorticity coupling  present in Eq.(\ref{eos1}) can increase the rate of cooling of the fireball. This may lead to early hadronization. Furthermore, we investigate its consequences on the thermal dilepton production from plasma. As anticipated from the effect of vorticity on temperature, the dilepton yield is suppressed and the suppression is more for the increased value of the initial vorticity. 

Our paper is organized as follows, in section[\ref{vorticityevolution}] we calculate the vorticity evolution by employing the hydrodynamic analysis. This is followed by a discussion on spin polarization tensor and its effect on the standard thermodynamic relation in section[\ref{spintensor}]. In section[\ref{expansion}], we study the temperature evolution of the fireball and discuss the effect of local vorticity on critical temperature. In section[\ref{dilepton}], we describe the vorticity effect on the thermal dilepton production. Finally, in section[\ref{conclusion}], we summarize and draw the conclusion.
\section{Vorticity evolution in HIC}
\label{vorticityevolution}
\begin{figure}[tbh]
\centering
\includegraphics[width=7.5cm]{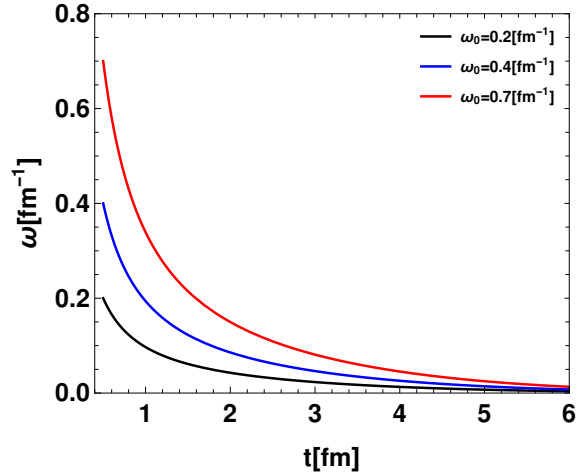}
\caption{Variation of vorticity ($\omega$) as a function of time for different values of initial vorticity ($\omega_0$) at $t_0=0.5$ fm.}
\label{vort}
\end{figure}

In this section, we discuss the time evolution of vorticity in QGP by employing  hydrodynamic analysis. 
The relativistic Euler equation which for an ideal fluid can be written as
\begin{equation}
(\epsilon+P)u^{\mu}\partial_{\mu}u^{\alpha}=\nabla^{\mu \alpha}\partial_{\mu}P,
\label{euler}
\end{equation} 
where, $\epsilon$ and $P$ are energy and pressure density respectively. Here, $u^{\mu}=\gamma(1,\vec{v})$ is fluid four-velocity with $\gamma^{-1}=\sqrt{1-v^2}$ and $\nabla^{\mu \alpha}=g^{\mu \alpha}-u^{\mu}u^{\alpha}$. Using the thermodynamic relation $P(\epsilon+P)^{-1}=c_s^2 \ln s$ and the second law of thermodynamics ($\partial_{\mu}s^{\mu}=0$), the spatial component of Eq.(\ref{euler}) can be written as
\begin{equation}
\gamma \frac{\partial }{\partial t}(\gamma \vec{v})+\gamma (\vec{v}\cdot\vec{\nabla})\gamma \vec{v}=-c_s^2\vec{\nabla}s+c_s^2\gamma \vec{v}\partial_{\mu}u^{\mu},
\label{vorevol}
\end{equation}
where $s$ is entropy density and $c_s$ is the speed of sound. Here, we work in the non-relativistic approximation and take $\gamma \approx 1$. This is justified when the fluid velocities are non-supersonic. Even for the fluid velocity of the order of the  sound speed $c_s=1/\sqrt{3}$ (for an ideal EoS $\epsilon=3p$), the value of $\gamma \equiv 1.2$. 
In the present work, we assume the flow to be subsonic and consider
the velocity to be non-relativistic and use equation of state $\epsilon=3p$ for massless constituents. Note that by taking this approximation we restrict ourselves in a region where velocity is small and the following definition of vorticity can be used~\cite{Deng:2016gyh}: 

\begin{equation}
\vec{\omega}(\vec{r})= \vec{\nabla} \times \vec{v}(\vec{r})
\label{vor}
\end{equation}
where $\vec{v}(r)$ is the fluid velocity. A similar approximation for vorticity has been considered in Ref.\cite{Deng:2016gyh,Csernai:2013bqa}.  In Ref.\cite{Csernai:2013bqa} the authors have studied vorticity development and distribution in both relativistic and non-relativistic limits by implementing numerical simulations and found that in both the limits the average values of  vorticity are of similar order. In a non-central collision, the average vorticity is perpendicular to the reaction plane. Taking $\hat{z}$ axis as the beam direction, $\hat{x}$ axis as impact parameter axis, the reaction plane is in the $xz$ plane, the average vorticity is established along $\hat{y}$ direction. For flow velocity, we adopt similar assumptions as considered in  Ref.(\cite{Deng:2016gyh}), where the fluid velocity $\vec{v}$ is decomposed into two parts; one part which is non-rotational flow ($\vec{v}_0$) and other being the rotational flow ($\vec{v}_r$) so that
\begin{equation}
\vec{v}=\vec{v}_{0}+\vec{v}_r.
\label{velocity}
\end{equation} 
The rotational flow velocity is defined in terms of vorticity as
\begin{equation}
\vec{v}_r=\frac{1}{2} \vec{\omega} \times \vec{r}.
\label{rotationalv}
\end{equation}
The non-rotational part $\vec{v}_0$ has three components, one is the longitudinal Bjorken flow part $v_{0z}=\frac{z}{t}$ and the other two components which come from transverse expansion are given by the differential equation ~\cite{Ollitrault:2008zz}
\begin{equation}
\frac{\partial \vec{v}_{0T}}{\partial t}=-\frac{\nabla_{T} P}{\epsilon+P}=-c_s^2 \nabla_{T} \ln(s),
\label{tflow}
\end{equation}
where $s$, $P$, $\epsilon$ respectively are the entropy, the pressure and the energy density. $\nabla_T$ is the gradient in the transverse direction ($x,y$). Following Ref.\cite{Ollitrault:2008zz}, we take the initial entropy density as
\begin{equation}
s \propto e^{-\frac{1}{2}\bigg(\frac{x^2}{\sigma_x^2}+\frac{y^2}{\sigma_y^2}+\frac{\eta_s^2}{\sigma_{\eta_s}^2}\bigg)},
\label{entropy}
\end{equation}
where $\sigma_x$ and $\sigma_y$ are transverse distribution root mean square widths ~\cite{Ollitrault:2008zz}. Solving Eq.(\ref{tflow}) for the transverse velocities by using the entropy density defined in Eq.(\ref{entropy}), we obtain
\begin{equation}
v_{0x}=\frac{c_s^2 x t}{\sigma_x^2},
\label{vx}
\end{equation}
\begin{equation}
v_{0y}=\frac{c_s^2 y t}{\sigma_y^2},
\label{vy}
\end{equation}
where $x$ and $y$ denote the position of the fluid cell and they can be obtained by solving the above two equations. The transverse expansion is due to the very high initial pressure gradient as noted in Ref.\cite{Ollitrault:2008zz} and dominate only in the time scale $t \gg \frac{\sigma_x}{c_s}$. For the vorticity evolution, taking curl of Eq.(\ref{vorevol}) and using Eqs.(\ref{velocity}), (\ref{rotationalv}), one obtain
\begin{equation}
\frac{d \vec{\omega}}{d t}=\vec{\nabla}\times(\vec{v}_{0}\times\vec{\omega}). \label{voreqn}
\end{equation}
Substituting in the above equation for $\vec{v}_0$, one can obtain the following solution for time dependence of $\vec{\omega}$ as in Ref.(\cite{Deng:2016gyh})
\begin{equation}
\vec{\omega}(\vec{r},t)=\frac{\omega_{0}{(\vec{r}_0,t_0)} t_0}{t}e^{-\frac{c_s^2}{2 \sigma_y^2}(t^2-t_{0}^2)}\hat{y},
\label{vorticity}
\end{equation}
where $\omega_{0}(\vec{r}_{0},t_0)$ is the initial vorticity at position $r_0$ and time $t_0$.  The factor $\frac{t_0}{t} e^{-\frac{c_s^2}{2 \sigma_y^2}(t^2-t_{0}^2)}$ represents the area swapped by the streamline from time $t_0$ to $t$. From the above equation it is clear that vorticity gets diluted due to the expansion. In Fig.(\ref{vort}), vorticity, as given by Eq.(\ref{vorticity}) is plotted as a function of time for different values of initial vorticities ($\omega_0$ and $t_0=0.5$ fm. Note that, with the $\omega_{0}$ values as taken here, the values for vorticities at a later time as evolved through Eq.(\ref{vorticity}) are of the similar order as reported by STAR\cite{STAR:2017ckg} at freeze out.

\section{Spin-polarization and Thermodynamic relation}
\label{spintensor}
 In this section, we discuss the spin polarization tensor and subsequent modification of the EoS. We start with the local thermal equilibrium distribution function for spin-1/2 particles which as a function of space and time can be written as~\cite{Becattini:2013fla}
\begin{equation}
f^{+}_{rs}(x,p)=\frac{1}{2 m} \bar{u}_{r}(p) X^{+} u_{s}(p)
\end{equation} 
\begin{equation}
f^{-}_{rs}(x,p)=-\frac{1}{2 m} \bar{v}_{s}(p) X^{-} v_{r}(p)
\end{equation} 
where $m$ is mass of the particle. $u_{r}(p)$ and $v_{s}(p)$ are bispinors with spin indices $r$ and $s$ running from 1 to 2. $X^{\pm}$, appearing in the above equations for the single particle distribution function is defined as the product of Boltzmann distribution function and the matrices $M^{\pm}$, can be written as~\cite{Becattini:2013fla,Florkowski:2017ruc,Florkowski:2018fap}
\begin{equation}
X^{\pm}=\exp[-\beta_{\mu}(x)p^{\mu}]M^{\pm} 
\label{x}
\end{equation}
where the matrices are defined as
\begin{equation}
M^{\pm}=\exp[\pm \frac{1}{2}{\omega}_{\mu \nu}\Sigma^{\mu \nu}].
\label{dist}
\end{equation}
In Eq.[\ref{x}], $\beta_{\mu}=\beta u^{\mu}$ with $\beta$ as inverse temperature ($\beta=1/T$) and $u^{\mu}=\gamma(1,\vec{v})$ is the flow velocity. In the definition of $M^{\pm}$,  ${\omega_{\mu \nu}}$ is the spin polarization tensor and $\Sigma_{\mu \nu}=\frac{i}{4}[\gamma_{\mu}, \gamma_{\nu}]$ is the spin operator in terms of Dirac matrices.

For the polarization tensor, we follow the tensor decomposition as done in Ref.~\cite{Florkowski:2017ruc} and write the anti-symmetric spin polarization tensor as 
\begin{equation}
\omega_{\mu \nu}=k_{\mu} u_{\nu}-k_{\nu}u_{\mu}+\epsilon_{\mu \nu \alpha \beta}u^{\alpha}\omega^{\beta}.
\label{pola}
\end{equation}
Here, $k_{\mu}$ and $\omega_{\mu}$ are orthogonal to the flow velocity $u_{\mu}$, so satisfies the relation $k^{\mu}u_{\mu}=\omega^{\mu}u_{\mu}=0$ and $\epsilon_{\mu \nu \alpha \beta}$ is the Levi-Civita tensor. For Levi-Civita and metric we use the following conventions: $\epsilon^{0123}=-\epsilon_{0123}=1$ and $g_{\mu \nu}=diag(1,-1,-1,-1)$. The quantities $k_{\mu}$ and $\omega_{\mu}$ can be written in terms spin polarization tensor as
\begin{equation}
k_{\mu}=\omega_{\mu \nu}u^{\nu}, \hspace{0.5cm} \omega_{\mu}=\frac{1}{2}\epsilon_{\mu \nu \alpha \beta}\omega^{\nu \alpha}u^{\beta}.
\label{kao}
\end{equation}
Here, we take the space like component of $\omega_{\mu}$ to be $\vec{\omega}=\vec{\nabla} \times\vec{v}$ as given in Eq.[\ref{vorticity}] together with time like component to be zero. Choosing $\omega^{\mu}=(0,0,\omega,0)$ i.e., rotation on $xz$ plane, one can solve Eqs.[\ref{pola}] and [\ref{kao}]  self consistently to construct the spin polarization tensor ($\omega_{\mu \nu}$) 
\begin{equation}
{\omega}_{\mu \nu}=
\begin{pmatrix}
0 & 0 & 0 & 0\\
0 & 0 & 0 & \frac{\omega}{T} \\
0 & 0 & 0 & 0 \\
0 & -\frac{\omega}{T} & 0 & 0
\end{pmatrix}
\label{thermal}
\end{equation}
The exponential in Eq.[\ref{dist}] is defined in terms of a power series and can be resummed, which, in general, can be complex ~\cite{Florkowski:2018fap}. As in Ref.~\cite{Florkowski:2017ruc}, we shall assume a special case when $\tilde{{\omega}}_{\mu \nu}\omega^{\mu \nu}=4 k\cdot \omega=0$, where  $\tilde{\omega}_{\mu \nu}$ is the dual of the polarization tensor i.e.,  $\tilde{\omega}_{\mu \nu}=\frac{1}{2} \epsilon_{\mu \nu \alpha \beta} \omega^{\alpha \beta}$, so that the matrix becomes
\begin{equation}
M^{\pm}=\cosh(\zeta)\pm \frac{\sinh(\zeta)}{2\zeta}\omega_{\mu \nu}\Sigma^{\mu \nu}
\end{equation} 
where $\zeta$ is given by
\begin{equation}
\zeta=\frac{1}{2 \sqrt{2}}\sqrt{\omega_{\mu \nu}\omega^{\mu \nu}}.
\label{zeta}
\end{equation}
Using Eqs.[\ref{thermal}]  [\ref{zeta}], \& $\zeta=\frac{\omega}{2 T}$ and 
 one can rewrite [\ref{eos1}] for $\mu =0 $:
\begin{equation}
\epsilon +P = Ts+ \frac{\omega}{2} w,
\label{eos2}
\end{equation}
where, $w=4 \cosh(\zeta) n_{0}$ with $n_0=\frac{T^3}{\pi^2}$ being number density of particles in the massless limit. The thermodynamic relation written in Eq.[\ref{eos2}] will be used for the temperature evolution and for calculation of the thermal dilepton production rate from a hot medium.

\section{Temperature evolution with finite vorticity}
\label{expansion}
Let us consider first the effect of vorticity on the profile of  4-velocity $u^{\mu}=(\gamma,\gamma v_x,0,\gamma v_z)$ where Lorentz factor $\gamma=\frac{1}{\sqrt{1-v^2}}$ with $v^2=v_x^2+v_z^2$. Here fluid velocity satisfy the  condition $u^{\mu}u_{\mu}=1$. As discussed in section \ref{vorticityevolution}, vorticity is along $\hat{y}$ direction so $v_y=0$.  $v_z$ and $v_x$ are respectively the longitudinal and the transverse components of velocity. Using Eq.[\ref{rotationalv}], along with longitudinal expansion these can be written as
\begin{equation}
v_x=\frac{1}{2} \omega z,
\end{equation}
\begin{equation}
v_z=\frac{z}{\tau}-\frac{1}{2}\omega x,
\end{equation}
where $x$ and $z$ are the positions coordinates.  $v_x$ is induced by vorticity and vanishes in the limit of non-vortical fluids. Similarly, for $\omega=0,$ the longitudinal velocity ($v_z$) becomes identical to the velocity in the Bjorken flow.  The energy-momentum tensor of the fluid can be written as
\begin{equation}
T^{\mu \nu}=(\epsilon+P)g^{\mu \nu}-P u^{\mu}u^{\nu},
\end{equation}
where, $\epsilon$ and $P$ are energy and pressure densities respectively. The temperature evolution equation can be obtained by using $\partial_{\mu}T^{\mu \nu}=0$ so that one can write:
 \begin{equation}
 u_{\nu}\partial_{\mu}T^{\mu \nu}=0,
\end{equation}
leading to
\begin{equation}
\partial_{\mu}((\epsilon+P)u^{\mu})=u_{\nu}g^{\mu \nu}\partial_{\mu}P.
\label{evol}
\end{equation}
Next, using  $\frac{d}{d\tau}=u^{\mu}\partial_{\mu}$ and standard thermodynamic relation $\epsilon+P=T s$, $c_s^2=\frac{d \epsilon}{d P}$, Eq.(\ref{evol}) can be rewritten in terms of proper time evolution of temperature as
\begin{equation}
\frac{dT}{d\tau}=-c_s^2 T \partial_{\mu}u^{\mu}.
\label{evolution}
\end{equation}
\noindent
In the case of 1-D Bjorken flow with $v_z=\frac{z}{\tau}$, the rhs in the above equation reduces to $-\frac{c_s^2 T}{\tau}$. In the presence of non-vanishing vorticity, the divergence of velocity profile ($\partial_{\mu} u^{\mu}$) will no longer be $\frac{1}{\tau}$ but contributions from vorticity dependent term in the velocity profile will also be present. However, this extra contribution will arise only from the Lorentz factor and will have a negligible contribution on temperature evolution. 
On the other hand, if one uses the modified thermodynamic relation Eq.[\ref{eos2}] with spin-vorticity coupling, 
one now gets the different form of temperature evolution:
\begin{equation}
\frac{dT}{d\tau}=-c_s^2 \bigg(T+\frac{\omega w}{2 s}\bigg) (\partial_{\mu} u^{\mu}).
\label{evolution1}
\end{equation} 
\begin{figure}[tbh]
\centering
\includegraphics[width=7.5cm]{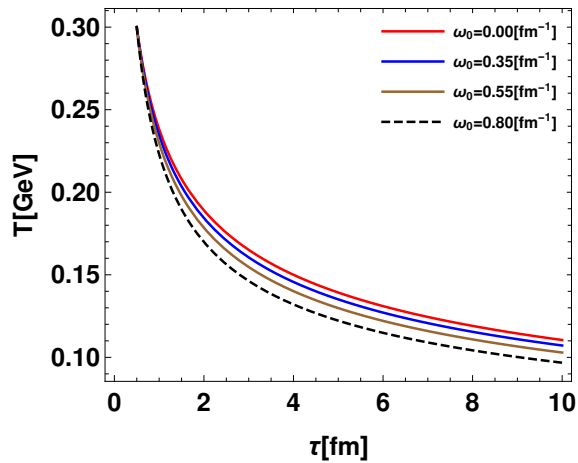}
\caption{
Variation of temperature with $\tau$ for various values of $\omega_0$ with $\tau_{0}=0.5$ fm and $T_{0}=300$ MeV. Red curve corresponds to usual 1-D Bjorken flow. With increase in vorticity the system cools faster.}
\label{flow}
\end{figure}
Using the velocity profiles as defined in Eqs.(\ref{vx}) and (\ref{vy}), we numerically solve temperature evolution equation i.e., Eq.(\ref{evolution1}). For this purpose, we take $c_s^2=\frac{1}{3}$, $\tau_0=0.5$ fm and $T_0=300$ MeV. For a non-central collision of $b=7$ fm, the rms widths are $\sigma_x=2$ fm, $\sigma_y=2.6$ fm. We take the vorticity profile as given in Eq.(\ref{vorticity}), where $\omega_0$ is a free parameter. Since the rotational motion should be such that the velocity should not exceed the speed of light, we consider $\omega_0 $ values so that the condition $\omega R < 1$ is always satisfied during the lifetime of the fireball. Figure(\ref{flow}) shows the behavior of temperature ($T$) vs time($\tau$). The red curve is the 1D Bjorken flow which we have reproduced in the limit of zero vorticity. As may be observed from Fig.[\ref{flow}], the spin-vorticity coupling term leads to faster cooling of the fireball as compared to the case with $\omega=0$. This also leads to a reduction of the hadronization time as shown in table(\ref{table1}). Here we have shown the results for initial temperature $T_0=300$ MeV and have taken critical temperature $T_c=150$ MeV. With the increase in vorticity, the critical time (by which system reaches at critical temperature) decreases.  
\begin{table}[htb]
\centering
\begin{tabular}{|c|c|c|c|}
\hline
 $T_0$(MeV) & $T_c$ (MeV)&$\omega_0$ (fm$^{-1}$)&$\tau_c$ (fm) \\ 
\hline
&  & 0.0 & 4.05   \\
&  & 0.2  & 3.90   \\
&  & 0.4  & 3.79   \\
300.0& 150.0 & 0.6  & 3.48  \\
&  & 0.8  &  3.10 \\
&  & 0.9  &  2.75 \\

\hline
\end{tabular}
\caption{Critical time $\tau_c$ for different values of $\omega_0$ with $\tau_{0}=0.5$ fm.}
\label{table1}
\end{table}
\section{Dilepton production}
\label{dilepton}
We next consider the effect of vorticity on the thermal dilepton production from QGP. The dominant channel here is the annihilation of quark and its anti-quark i.e., $q \bar{q}\rightarrow \gamma* \rightarrow l\bar{l}$. In the limit of massless quarks, dilepton production per unit space-time volume is written as \cite{Bhatt:2011kx}
\begin{equation}
\frac{d N}{d^4 x}=g^2 M^2\int \frac{d^3 p_1}{(2 \pi)^3}\frac{d^3 p_2}{(2 \pi)^3}\frac{f(E_1) f(E_2)}{2 E_1 E_2} \sigma(M),
\label{rate}
\end{equation}
where $p_{1,2}, E_{1,2}$ are lepton/anti-lepton momenta and energy. $f(E_i)^{-1}=1+e^{{\beta E_i}}$ is the distribution function of quark/anti-quark, $\sigma(M)$ is a cross-section of thermal dilepton production and $M$ is the invariant mass.  Let us note here that effect of the spin-vorticity coupling in Eq.[\ref{rate}] enters  through the temperature evolution. In Born approximation, $\sigma(M)=\sum_f\frac{16 \pi \alpha^2 q_f^2 N_c}{3 M^2 g^2}$, which for $N_f=2$ and $N_c=3$ reduces to  \cite{Alam:1996fd}
\begin{equation}
\sigma(M)=\frac{80 \pi \alpha^2}{9 M^2 g^2}.
\end{equation}
To determine the dilepton production rate for a given invariant mass $M$ and momentum $p$, we multiply Eq.[\ref{rate}] by unity by including two delta functions, one for invariant mass and other for momentum
\begin{equation}
 \int dM^2 \delta(M^2-s) d^3p \delta^3(p-p_1-p_2)=1
\end{equation}
where $p=(E_1+E_2,\vec{p_1}+\vec{p_2})$ is 4-momentum exchange of dileptons and $s$ is Mandelstan variable. By introducing the unity factor the dilepton production rate is written as
\begin{equation}
\frac{dN}{d^4x d^3p dM^2}=g^2 M^2\int \frac{d^3 p_1}{(2 \pi)^3}\frac{d^3 p_2}{(2 \pi)^3}\frac{f(E_1) f(E_2)}{2 E_1 E_2} \sigma(M)\delta(M^2-s) \delta^3(\vec{p}-\vec{p}_1-\vec{p}_2).
\label{drate}
\end{equation}
In the limit  $M\gg T$, Fermi-Dirac distribution function is replaced by Maxwell-Boltzmann distribution function and Eq.(\ref{drate}) reduces to
\begin{equation}
E\frac{dN}{d^4x d^3p dM^2}=\frac{1}{4}\frac{M^2 g^2 \sigma(M)}{(2 \pi)^5}e^{-\frac{p_0}{T}}
\label{drate1}
\end{equation}
where $E=E_1+E_2$ is the energy of the lepton pair. Let us note that Eq.(\ref{drate1}) is in the rest frame of the system, for an expanding system $p_0$ is replaced by ${u} \cdot {p}=u^{\mu}p_{\mu}$, where $u_{\mu}$ is fluid four-velocity and
\begin{equation}
p^{\mu}=(m_T \cosh y,p_T \cos \phi,p_T \sin \phi,m_T \sinh y).
\end{equation}
\begin{figure}[tbh]
\subfigure{
\hspace{-0mm}\includegraphics[width=6.35cm]{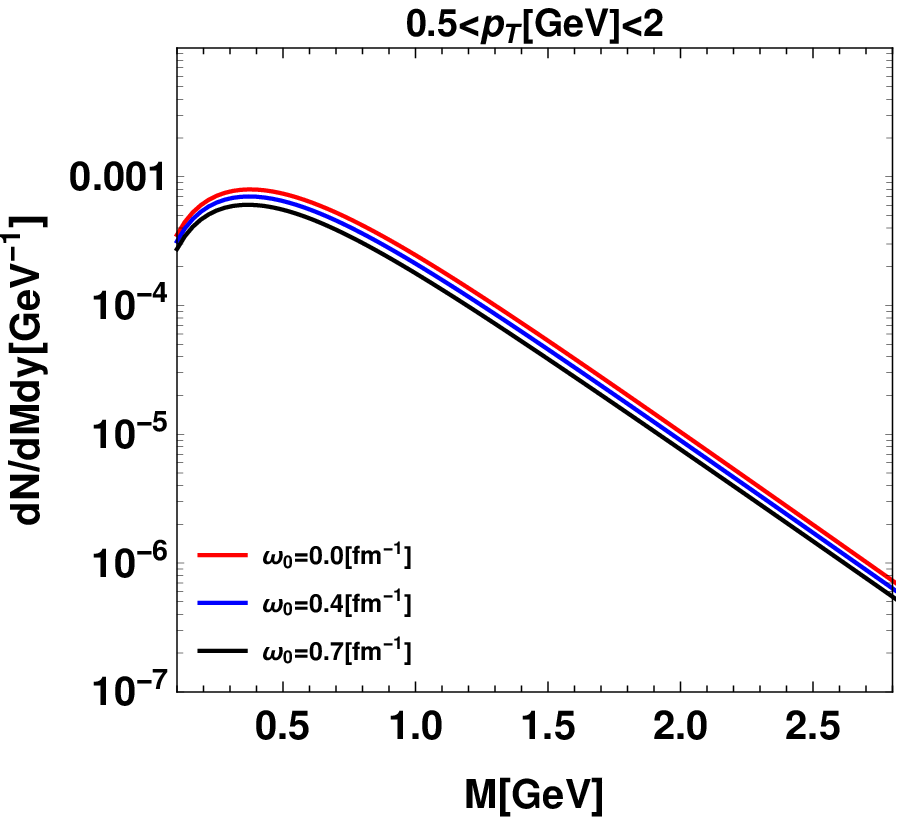}} 
\subfigure{
\includegraphics[width=6.35cm]{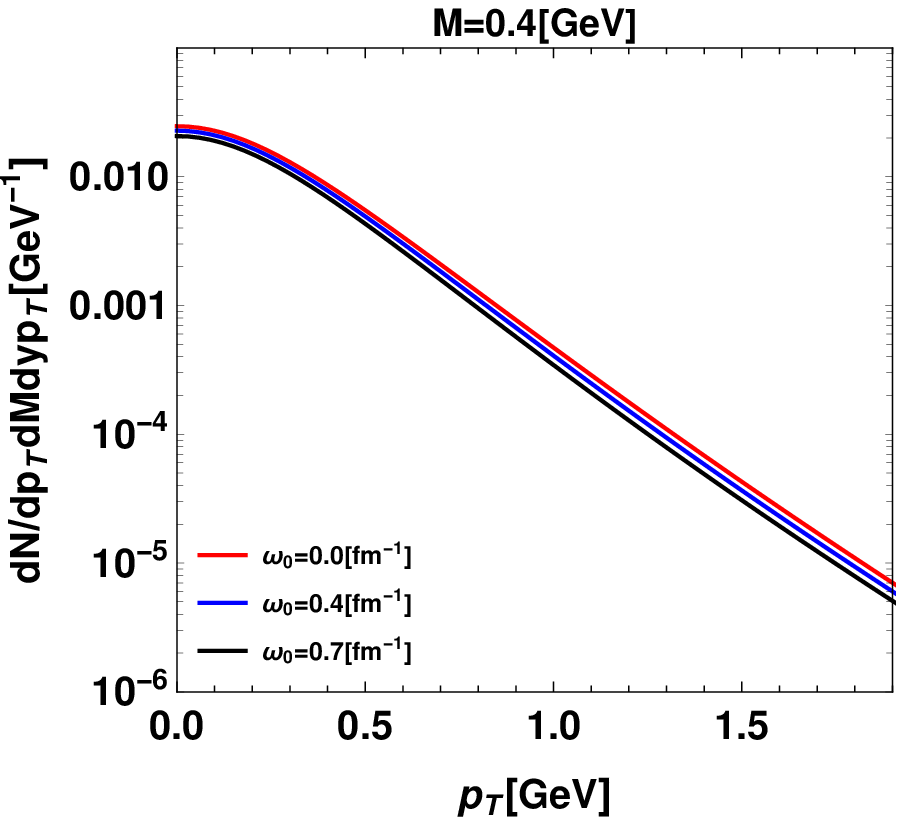}}
\caption{
Left: Dilepton production as a function of invariant mass for different values of $\omega_0$, $\tau_{0}=0.5$ fm and $p_{T}=(0.5-2)$ GeV. Right: Dilepton production as a function of transverse momentum for different values of $\omega_0$, $M=0.4$ GeV and $\tau_{0}=0.5$ fm.}
\label{rate1}
\end{figure}
Here, $\eta_s$ is space-time rapidity, $p_T$ is particle's transverse momentum, $\phi$ is azimuthal angle and $m_T=\sqrt{p_T^2+M^2}$. For an expanding system fluid four velocity $u^{\mu}$ can be written as
\begin{equation}
u^{\mu}=(\cosh\eta_s \cosh \xi, \sinh \xi, 0, \sinh \eta_s \cosh \xi),
\label{flowvel}
\end{equation}
where, $\sinh \xi=\gamma v_{x}$ and $\tanh(\eta_s)=v_z$. In the space-time volume element $d^4x=dtdzd^2x$, $d^2x$ is transverse area and $dt,dz$ are lengths elements along time and longitudinal directions. Using the parameterization $t=\tau \cosh\eta_s \cosh \xi$, $x=\tau \sinh \xi$, $z=\tau \sinh \eta_s \cosh \xi$, where $\tau=\sqrt{t^2-x^2-z^2}$ and $\tanh\eta_s=\frac{z}{t}$, and the transverse area $d^2x=r dr d\phi$, the space-time volume $d^4x$ can be written as 
\begin{equation}
d^4x=\frac{1}{2}\tau R^2 d\phi d\tau d\eta_s,
\end{equation}
where $\phi$ is azimuthal angle and $R=1.2 (A)^{\frac{1}{3}}$ is the nucleus radius used for collision (for Gold, A=197). We will calculate differential rates as a function of invariant mass and transverse momentum. These production rates as a function of dilepton invariant mass ($M$), transverse momentum $p_T$ and particle rapidity $y$ can be obtained by using $\frac{d^3p}{E}=2 \pi p_T dp_T dy$ together with Eq.[\ref{drate1}] and are written as \cite{florkowski}
\begin{equation}
\frac{dN}{dM dy}=4 \pi M R^2\int_{\tau_0}^{\tau}\tau d\tau \int_{0}^{2 \pi}d\phi \int_{-\eta_{min}}^{\eta_{max}}d \eta_s \int p_T dp_T\bigg(E\frac{dN}{d^4x d^3p dM^2}\bigg),
\label{ratem}
\end{equation} 
and 
\begin{equation}
\frac{dN}{p_T dp_T dM dy}=4 \pi M R^2\int_{\tau_0}^{\tau}\tau d\tau \int_{0}^{2 \pi}d\phi \int_{-\eta_{min}}^{\eta_{max}}d \eta_s \bigg(E\frac{dN}{d^4x d^3p dM^2}\bigg),
\label{ratept}
\end{equation}
where $\tau_0$ and $\tau$ respectively are the initial thermalization time and the time by which the temperature of the system reaches at its critical value ($T_c$). 
\begin{figure}[tbh]
\subfigure{
\hspace{-0mm}\includegraphics[width=6.35cm]{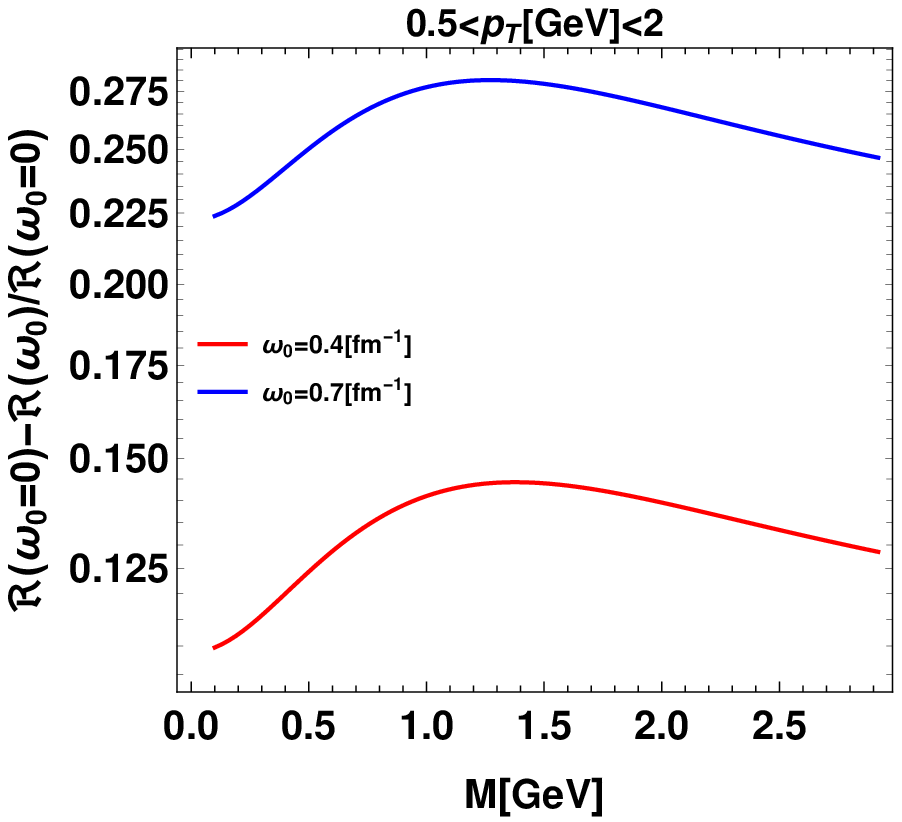}} 
\subfigure{
\includegraphics[width=6.35cm]{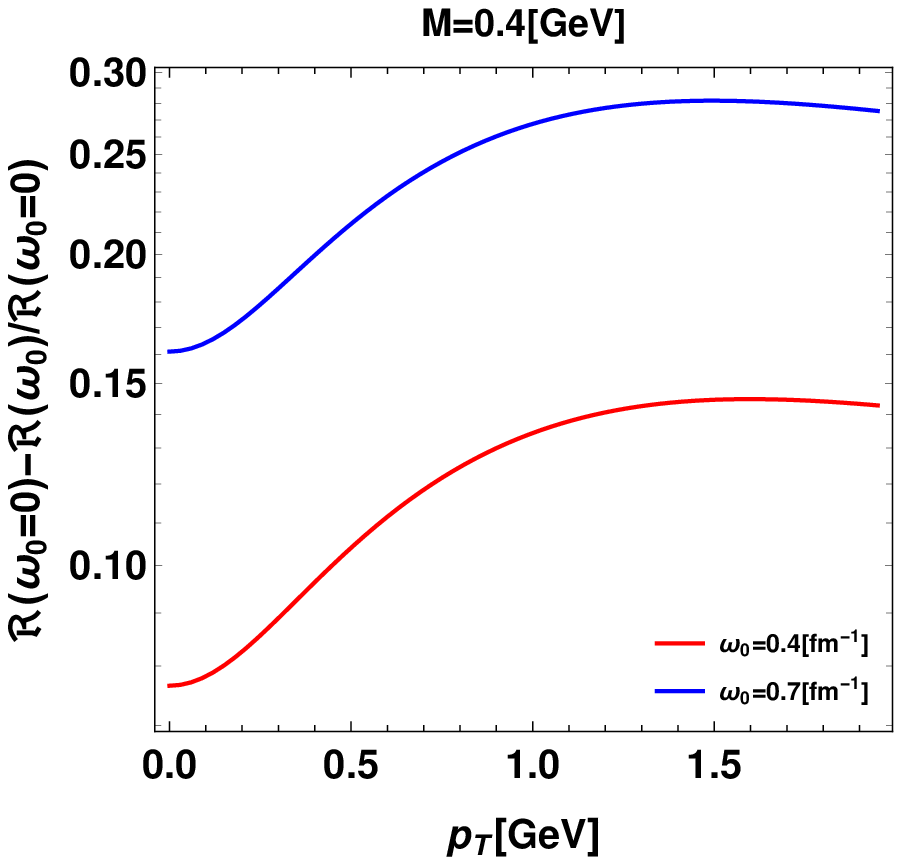}}
\caption{
Left: Fractional change in dilepton production as a function of invariant mass for different values of $\omega_0$. Right: Fraction change in dilepton production as a function of transverse momentum for different values of $\omega_0$, $M=0.4$ GeV.}
\label{ratio}
\end{figure}
Here we have taken $\tau_0=0.5$ fm and critical temperature $T_c=150 $ MeV. $\tau$ for different values of $\omega_0$ are shown in table \ref{table1}. Further, we take $\eta_{max}= 5.3$ and the particle rapidity $y=0$. 

The results for differential production rate as a function of invariant mass (left) and as a function of transverse momentum (right) are shown in Fig(\ref{rate1}). As anticipated from the cooling rate of the plasma the dilepton production rate is suppressed in the presence of vorticity. The red curve is the production rate without vorticity which is consistent with the results shown in Ref\cite{Rapp:2000pe}. In Fig.(\ref{ratio}), the fractional change in dilepton production due to vorticity is shown. Here $\mathcal{R}$ denotes corresponding production rates as defined in Eqs.(\ref{ratem}) and (\ref{ratept}). Left figure is variation with the invariant mass and the right one is with the transverse momentum ($p_T$). In the left figure, it is clear that suppression is maximum at an invariant mass around $1$ GeV. For an initial vorticity $\omega_0=0.4$ fm$^{-1}$ maximum suppression is 15\% and for $\omega_0=0.7$ fm$^{-1}$ the suppression can be as large as 28\% as compared to the case of zero vorticity. On the other hand for $p_T$ variation (right figure) suppression is more around transverse momentum $p_T=1-1.5$ GeV. For the same values of initial vortices $\omega_0=0.4$ fm$^{-1}$ and $\omega_0=0.7$ fm$^{-1}$ the maximum suppression is about 15\% and 28\% respectively.  For $p_T$ less than $1$ GeV the suppression is small.
\section{Summary and Conclusion}
\label{conclusion}

In this work, we have analyzed the role of spin-polarization and vorticity on the evolution of the QGP
created in relativistic heavy-ion collisions. Because of the initial vorticity, one needs to modify the Bjorken flow describing the initial stages of hydrodynamic evolution. Inclusion of vorticity leads to leading to a 2+1 dimensional hydrodynamic expansion of the system. We find that in the absence of spin-polarization, vorticity
alone cannot significantly influence the temperature evolution of the QGP. This situation changes when the effect
of spin-vorticity coupling is incorporated in the thermodynamic relation given in Eq.(\ref{eos2}).
Our analysis shows that the expanding plasma cools at a much faster rate in comparison with the case without 
the spin-polarization. This can lead to early hadronization of the system as shown in Table (\ref{table1}). 
Further, we have studied thermal dilepton production in  this scenario and found that the production rates
are suppressed due to the faster cooling of the system. 
We emphasize on the fact that the analytic solutions used here for velocity profile may not be valid at late time evolution. Within this limit, we predict the suppression of dilepton production and early hadronization as a consequence of spin vorticity coupling. Our results are also useful in testing the presence of a vorticity induced term in the thermodynamic relation i.e., Eq.(\ref{eos2}). If such a term is present in the thermodynamic relation then we predict that its effect can be studied via the production of thermal dilepton pairs. This is a first attempt to include vorticity in the study of thermal evolution and thermal dilepton production.


\begin{thebibliography}{99}
\bibitem{Gale:2013da} 
C.~Gale, S.~Jeon and B.~Schenke,
Int.\ J.\ Mod.\ Phys.\ A {\bf 28}, 1340011 (2013)
%
\bibitem{Heinz:2013th} 
U.~Heinz and R.~Snellings,
Ann.\ Rev.\ Nucl.\ Part.\ Sci.\  {\bf 63}, 123 (2013)
%
\bibitem{Kovtun:2004de} 
P.~Kovtun, D.~T.~Son and A.~O.~Starinets,
Phys.\ Rev.\ Lett.\  {\bf 94}, 111601 (2005)
\bibitem{Becattini:2007sr} 
F.~Becattini, F.~Piccinini and J.~Rizzo,
Phys.\ Rev.\ C {\bf 77}, 024906 (2008)
\bibitem{STAR:2017ckg} 
L.~Adamczyk {\it et al.} [STAR Collaboration],
Nature {\bf 548}, 62 (2017)
%
\bibitem{Jiang:2016woz} 
Y.~Jiang, Z.~W.~Lin and J.~Liao,
Phys.\ Rev.\ C {\bf 94}, no. 4, 044910 (2016)
Erratum: [Phys.\ Rev.\ C {\bf 95}, no. 4, 049904 (2017)]
%
\bibitem{Gao:2014coa} 
J.~H.~Gao, B.~Qi and S.~Y.~Wang,
Phys.\ Rev.\ D {\bf 90}, no. 8, 083001 (2014)
%
\bibitem{Csernai:2014hva} 
L.~P.~Csernai, D.~J.~Wang and T.~Csorgo,
Phys.\ Rev.\ C {\bf 90}, no. 2, 024901 (2014)
%
\bibitem{Deng:2016gyh} 
W.~T.~Deng and X.~G.~Huang,
Phys.\ Rev.\ C {\bf 93}, no. 6, 064907 (2016)
%
\bibitem{Xie:2016fjj} 
Y.~L.~Xie, M.~Bleicher, H.~Stöcker, D.~J.~Wang and L.~P.~Csernai,
Phys.\ Rev.\ C {\bf 94}, no. 5, 054907 (2016)
%
\bibitem{Liang:2004ph} 
Z.~T.~Liang and X.~N.~Wang,
Phys.\ Rev.\ Lett.\  {\bf 94}, 102301 (2005)
Erratum: [Phys.\ Rev.\ Lett.\  {\bf 96}, 039901 (2006)]
%
\bibitem{Karpenko:2016jyx} 
I.~Karpenko and F.~Becattini,
Eur.\ Phys.\ J.\ C {\bf 77}, no. 4, 213 (2017)
%
\bibitem{Betz:2007kg} 
B.~Betz, M.~Gyulassy and G.~Torrieri,
Phys.\ Rev.\ C {\bf 76}, 044901 (2007)
%
\bibitem{Upsal:2016phr} 
I.~Upsal [STAR Collaboration],
J.\ Phys.\ Conf.\ Ser.\  {\bf 736}, no. 1, 012016 (2016).
%

\bibitem{Ipp:2007ng} 
A.~Ipp, A.~Di Piazza, J.~Evers and C.~H.~Keitel,
Phys.\ Lett.\ B {\bf 666}, 315 (2008)
%
\bibitem{Liang:2004xn} 
Z.~T.~Liang and X.~N.~Wang,
Phys.\ Lett.\ B {\bf 629}, 20 (2005)
%
\bibitem{Csernai:2013vda} 
L.~P.~Csernai, S.~Velle and D.~J.~Wang,
Phys.\ Rev.\ C {\bf 89}, no. 3, 034916 (2014)
%
\bibitem{Rogachevsky:2010ys} 
O.~Rogachevsky, A.~Sorin and O.~Teryaev,
Phys.\ Rev.\ C {\bf 82}, 054910 (2010)
%
\bibitem{Kharzeev:2007tn} 
D.~Kharzeev and A.~Zhitnitsky,
Nucl.\ Phys.\ A {\bf 797}, 67 (2007)
%
\bibitem{Jiang:2015cva} 
Y.~Jiang, X.~G.~Huang and J.~Liao,
Phys.\ Rev.\ D {\bf 92}, no. 7, 071501 (2015)
%
\bibitem{Bernett:2015}
S.J.Barnett, Phys.Rev.6, 239 (1915).
%
\bibitem{Fukushima:2018osn} 
K.~Fukushima, S.~Pu and Z.~Qiu,
arXiv:1808.08016 [hep-ph].
%
\bibitem{Becattini:2009wh} 
F.~Becattini and L.~Tinti,
Annals Phys.\  {\bf 325}, 1566 (2010)
%
\bibitem{Florkowski:2017ruc} 
W.~Florkowski, B.~Friman, A.~Jaiswal and E.~Speranza,
Phys.\ Rev.\ C {\bf 97}, no. 4, 041901 (2018)
%
\bibitem{Becattini:2013fla} 
F.~Becattini, V.~Chandra, L.~Del Zanna and E.~Grossi,
Annals Phys.\  {\bf 338}, 32 (2013)
%
\bibitem{Becattini:2018duy} 
F.~Becattini, W.~Florkowski and E.~Speranza,
Phys.\ Lett.\ B {\bf 789}, 419 (2019)
%
\bibitem{Florkowski:2018fap} 
W.~Florkowski and R.~Ryblewski,
arXiv:1811.04409 [nucl-th].
%
\bibitem{Csernai:2013bqa} 
L.~P.~Csernai, V.~K.~Magas and D.~J.~Wang,
Phys.\ Rev.\ C {\bf 87}, no. 3, 034906 (2013)
%
\bibitem{Ollitrault:2008zz} 
J.~Y.~Ollitrault,
Eur.\ J.\ Phys.\  {\bf 29}, 275 (2008)
%

\bibitem{Bhatt:2011kx} 
J.~R.~Bhatt, H.~Mishra and V.~Sreekanth,
Nucl.\ Phys.\ A {\bf 875}, 181 (2012)
%
\bibitem{Alam:1996fd} 
J.~Alam, B.~Sinha and S.~Raha,
Phys.\ Rept.\  {\bf 273}, 243 (1996).
%
\bibitem{florkowski} W. Florkowski,
Phenomenology
of Ultra-Relativistic Heavy-Ion Collisions
(World Scientific,
Singapore, 2010).
%
\bibitem{Rapp:2000pe} 
R.~Rapp,
Phys.\ Rev.\ C {\bf 63}, 054907 (2001).
\end{thebibliography}
\end{document}